\documentclass[aps,prl,reprint,superscriptaddress,showpacs]{revtex4-1}
\usepackage[latin9]{inputenc}
\usepackage{babel}
\usepackage{amsmath}
\usepackage{amssymb}
\usepackage{graphicx}
\usepackage[unicode=true,
 bookmarks=false,
 breaklinks=false,pdfborder={0 0 1},colorlinks=true,linkcolor=red,citecolor=blue]
 {hyperref}
\makeatletter
\@ifundefined{textcolor}{}
{%
 \definecolor{BLACK}{gray}{0}
 \definecolor{WHITE}{gray}{1} 
 \definecolor{RED}{rgb}{1,0,0}
 \definecolor{GREEN}{rgb}{0,1,0}
 \definecolor{BLUE}{rgb}{0,0,1}
 \definecolor{CYAN}{cmyk}{1,0,0,0}
 \definecolor{MAGENTA}{cmyk}{0,1,0,0}
 \definecolor{YELLOW}{cmyk}{0,0,1,0}
}

\usepackage{amsthm}
\usepackage{amsfonts}
\usepackage{bbm}
\usepackage{epsfig}
\usepackage{times}
\usepackage{babel}
\usepackage{color}
\usepackage{framed}
\usepackage{changes}
\usepackage{float}
\usepackage{array}

\makeatother

\begin{document}
\title{Critical phenomena in dynamical scalarization of charged black hole}
\author{Cheng-Yong Zhang}
\email{zhangcy@email.jnu.edu.cn}
\address{\textit{Department of Physics and Siyuan Laboratory, Jinan University,
Guangzhou 510632, China}}

\author{Qian Chen}
\email{chenqian192@mails.ucas.ac.cn}
\address{\textit{School of Physical Sciences, University of Chinese Academy
of Sciences, Beijing 100049, China}}

\author{Yunqi Liu}
\email{yunqiliu@yzu.edu.cn (corresponding author)}
\address{\textit{Center for Gravitation and Cosmology, College of Physical
Science and Technology, Yangzhou University, Yangzhou 225009, China}}

\author{Wen-Kun Luo}
\email{luowk@stu2020.jnu.edu.cn}
\address{\textit{Department of Physics and Siyuan Laboratory, Jinan University,
Guangzhou 510632, China}}

\author{Yu Tian}
\email{ytian@ucas.ac.cn}
\address{\textit{School of Physical Sciences, University of Chinese Academy
of Sciences, Beijing 100049, China}}
\address{\textit{Institute of Theoretical Physics, Chinese Academy of Sciences,
Beijing 100190, China}}

\author{Bin Wang}
\email{wang\_b@sjtu.edu.cn (corresponding author)}
\address{\textit{Center for Gravitation and Cosmology, College of Physical
Science and Technology, Yangzhou University, Yangzhou 225009, China}}
\address{\textit{Shanghai Frontier Science Center for Gravitational Wave Detection, Shanghai Jiao Tong
University, Shanghai 200240, China}}

\begin{abstract}
 
We report a new black hole (BH) scalarization mechanism and disclose novel dynamical critical phenomena in the process of the nonlinear accretion of the scalar field into BHs. The accretion process can transform a seed BH into a final scalarized or bald BH, depending on the initial parameter of the scalar field $p$.  There is a critical parameter $p_{\ast}$ and near it all intermediate solutions are attracted to a critical solution (CS) and stay there for a time scaling as $T\propto-\gamma\ln|p-p_{\ast}|$. 
At late times, the solutions evolve into scalarized black holes (BHs) if $p>p_{\ast}$, or bald BHs if $p<p_{\ast}$. The final masses of the resulting scalarized/bald BHs satisfy power-laws $M_{p}-M_{\pm}\propto|p-p_{\ast}|^{\gamma_{\pm}}$ where $M_{\pm}$ are the masses of the scalarized/bald BHs when $p\to p_\ast$ from above/below, and $\gamma_{\pm}$ the corresponding exponents.

\end{abstract}
\maketitle

\section{Introduction}

One of the most intriguing phenomena in BH physics is the critical behavior in gravitational collapse, which has highlighted  the nonlinear dynamics at the threshold of BH formation \citep{Choptuik:1992jv}.
Choptuik first observed in the study of a massless scalar field
collapse that there is generally a critical parameter value $p_{\ast}$ which signals the onset of BH formation for each family of initial data parameterized by $p$. 
In the subcritical regime ($p<p_{\ast}$), the scalar field disperses to infinity, leaving a flat spacetime behind. However in the supercritical regime ($p>p_{\ast}$), a BH forms with mass satisfying the power-law $M\propto(p-p_{\ast})^{\gamma}$ in which  $\gamma$ is a universal exponent. 
 The imploding scalar wave induces a phase transition between the flat/black-hole spacetimes. 
The BH turns on with an infinitesimal mass for the type II transition, where 
the CS at the threshold is self-similar and universal \citep{Abrahams:1993wa,Evans:1994pj,Choptuik:2004ha,Koike:1995jm,Gundlach:1995kd,Garfinkle:1998va}.  For the type I transition in some theories, the BH turns on with a finite mass near the threshold, but does not follow the power-law. Instead, the intermediate evolutions stay near the CS with a time scaling as  $T\propto-\gamma\ln|p-p_{\ast}|$. The universal CS is stationary or time-periodic \citep{Bartnik:1988am,Seidel:1991zh,Bizon:1998kq,Liebling:1996dx,Choptuik:1996yg,Brady:1997fj}. 
For a detailed review, please refer to \citep{Gundlach:2007gc}.

The critical behaviors reported were found between the no-black-hole/black-hole transition. 
In this work, 
we introduce novel dynamical critical phenomena during the nonlinear simulation of the bald/scalarized BH transition in theories with nonminimal coupling between the scalar and a source term.  
The scalarized BH can be formed by accretion of the scalar field surrounding a  seed BH at the center. For the initial scalar data parameterized by $p$, there is a threshold $p_\ast$ at where a metastable scalarized critical BH solution lives.  Near the threshold, all the intermediate solutions are attracted to this CS and stay there for a time scaling as  $T\propto-\gamma\ln|p-p_{\ast}|$. At late times, the intermediate solutions decay  
to bald BHs if $p<p_{\ast}$, or to scalarized BHs if $p>p_{\ast}$.  The final values of the scalar and BH mass are discontinuous across the critical point. Thus the critical scalarization is a kind of first-order phase transition. But unlike the case in type I critical gravitational collapse, we find that final masses of the scalarized/bald BHs follow the power-law $M_{p}-M_{\pm}\propto|p-p_{\ast}|^{\gamma_{\pm}}$ in which $M_{\pm}$ are the masses of the scalarized/bald BHs when $p\to p_\ast$ from above/below, and $\gamma_{\pm}$ the corresponding exponents.

The nonlinear accretion of the scalar field introduces a new mechanism for the BH scalarization, which is different from the spontaneous scalarization triggered by the linear tachyonic instability of the scalar field in bald BH background. Either the bald or scalarized BH is linearly stable at the same point in the model parameter space, and no intermediate attractor appears in the dynamical simulation for the spontaneous scalarization \citep{Zhang:2021etr}.  However, the new scalarization we uncovered reflects the consequences of nonlinearity. The CS behaves as an attractor, and both the final bald and scalarized BHs are linearly stable at the same point in the model parameter space. 

The specific action we consider is 
\begin{equation}
S=\frac{1}{16\pi}\int d^{4}x\sqrt{-g}\left[R-2\nabla_{\mu}\phi\nabla^{\mu}\phi-f(\phi)I\right],
\end{equation}
where $R$ is the Ricci scalar for the metric $g_{\mu\nu}$. The
real scalar field $\phi$ couples to a source term $I$ through function
$f(\phi)$. It is known as the Einstein-Maxwell-scalar (EMS) theory
if $I=F_{\mu\nu}F^{\mu\nu}$, or the Einstein-scalar-Gauss-Bonnet
(EsGB) theory if $I=R^{2}-4R_{\mu\nu}R^{\mu\nu}+R_{\mu\nu\rho\sigma}R^{\mu\nu\rho\sigma}$.
These theories have attracted many attentions due to the spontaneous
scalarization \citep{Doneva1711,Silva1711,Antoniou1711,Cunha1904,Dima:2020yac,Herdeiro2009,Berti2009, Herdeiro:2018wub}.
To obtain a theory allowing the bald  
solution, there should be $\frac{df}{d\phi}(0)=0$. The spontaneous
scalarization occurs if $\frac{d^{2}f}{d\phi^{2}}(0)>0$ for EMS
theory, or $\frac{d^{2}f}{d\phi^{2}}(0)<0$ for EsGB theory, due
to the fact that the effective mass squared of the scalar perturbation
is negative, and the tachyonic instability of the bald BH is triggered.
Here
we focus on the EMS theory with coupling function $f(\phi)=e^{\beta\phi^{4}}$,
in which $\beta$ is a  parameter. This model allows
the bald solution, but the spontaneous scalarization is quenched
since $\frac{d^{2}f}{d\phi^{2}}(0)=0$.
In other words, the bald Reissner-Nordstr\"om (RN)
BH in this theory is linearly stable under small perturbation. But we find it is nonlinearly unstable under large disturbance and evolves into a scalarized BH.
We confirm that the EMS theories with other coupling functions such as $f(\phi)=e^{\beta\phi^n},1+\beta\phi^n$ with $n=3,4,5,6$ also have {the nonlinear instability and the critical phenomena in dynamical scalarization}.

\section{Numerical setup}
We study the nonlinear dynamics of the spherically symmetrical BHs under large disturbance in EMS theory by adopting the Painlev\'e-Gullstrand-like coordinates ansatz
\begin{equation}
ds^{2}=-\left(1-\zeta^{2}\right)\alpha^{2}dt^{2}+2\zeta\alpha dtdr+dr^{2}+r^{2}d\Omega_{2}^{2}.\label{eq:PG}
\end{equation}
Here $d\Omega_{2}^{2}$ is the line element of unit sphere $S^{2}$
and $\alpha,\zeta$ are metric functions of $(t,r)$.
This coordinate system is regular on the apparent horizon which locates
at $\zeta=1$.  For a bald RN BH, $\alpha=1$ and $\zeta=\sqrt{\frac{2M}{r}-\frac{Q^{2}}{r^{2}}}.$

Taking the gauge potential as $A_{\mu}dx^{\mu}=A(t,r)dt$, the Maxwell
equations give
\begin{equation}
\partial_{r}A=\frac{Q\alpha}{r^{2}f(\phi)},\label{eq:Ar}
\end{equation}
in which $Q$ is the electric charge parameter. For scalarized solutions, the coupling between scalar and electromagnetic field can modulate electromagnetic energy and transform it into the scalar. 
Introducing auxiliary variables $\Phi=\partial_{r}\phi$
and $\Pi=\frac{1}{\alpha}\partial_{t}\phi-\zeta\Phi,$ the Einstein
equations give
\begin{align}
\partial_r\zeta= & \frac{r}{2\zeta}\left(\Phi^{2}+\Pi^{2}+\frac{Q^{2}}{r^{4}f}\right)-\frac{\zeta}{2r}+r\Pi\Phi,\label{eq:zr}\\
\partial_r\alpha= & -\frac{r\Pi\Phi\alpha}{\zeta},\label{eq:ar}\\
\partial_{t}\zeta= & \frac{r\alpha}{\zeta}\left(\Pi+\Phi\zeta\right)\left(\Pi\zeta+\Phi\right).\label{eq:zt}
\end{align}
The scalar equations can be written as
\begin{align}
\partial_{t}\phi & =\alpha\left(\Pi+\Phi\zeta\right),\label{eq:phit}\\
\partial_{t}\Pi & =\frac{\partial_{r}\left[\left(\Pi\zeta+\Phi\right)\alpha r^{2}\right]}{r^{2}}-\frac{\alpha Q^{2}}{2r^{4}f{}^{2}}\frac{df}{d\phi}.\label{eq:Pt}
\end{align}

Given initial $\phi$ and $\Pi$, we can get the initial $\Phi$ and
then $\zeta,\alpha$ from constraints (\ref{eq:zr}, \ref{eq:ar}).
The $\zeta,\phi,\Pi$ on next time slices can be obtained from
the evolution equations (\ref{eq:zt}, \ref{eq:phit}, \ref{eq:Pt}).
The constraint (\ref{eq:zr}) is used only once at the beginning.

At  spatial infinity, the matter functions $\phi,\Pi,\Phi$
should be zero. Then $\zeta\to\sqrt{\frac{2M}{r}}$ as $r\to\infty$.
Here the constant $M$ is the total Misner-Sharp mass of the spacetime.
We introduce $s=\sqrt{r}\zeta$ to replace $\zeta$ in the simulation
and set the boundary conditions for $s,\alpha$ as
\begin{equation}
s|_{r\to\infty}=\sqrt{2M},\ \ \alpha|_{r\to\infty}=1.\label{eq:boundary}
\end{equation}
The second equality implies that the coordinate time $t$ equals the proper time at the  infinity. The computational domain ranges in $(r_{0},\infty)$ where $r_{0}$ is a bit smaller
than the initial apparent horizon. We use the fourth-order finite
difference method in the radial direction by compactification $z=\frac{r}{r+M}$ and discretizing $z$ uniformly
with about $2^{11}\sim2^{12}$ grid points. The resolution is limited
in the far region at late times. But it is accurate enough since
we focus on the near horizon behavior. The time evolution is solved
with the fourth-order Runge-Kutta method. The Kreiss-Oliger dissipation
is employed to stabilize the simulation. At the first step, the constraint (\ref{eq:zr}) is solved by the Newton-Raphson method \cite{Dias:2015nua}.

We use the following families of initial data $\phi=ae^{-(\frac{r-cM}{wM})^{2}}$, $\Pi=0$. Here $a,c,w$ parameterize the initial amplitude, center, and width
of the Gaussian wave, respectively. Other types of initial data such
as $\phi=a\left(1-\tanh\frac{r}{wM}\right),\Pi=0$ are also employed. If the initial scalar locates far away from the center, one gets an initial RN BH at the center. If the initial scalar locates near the center, one gets an initial non-equilibrium charged BH, which deviates from the  RN BH. 
For all these initial data families, we observe qualitatively the same results.
We check the accuracy and convergence of our numerical method in
various ways and
find that it converges
to the fourth-order.

\section{Numerical results}

\begin{figure}[h]
    \begin{centering}
    \includegraphics[width=0.80\linewidth]{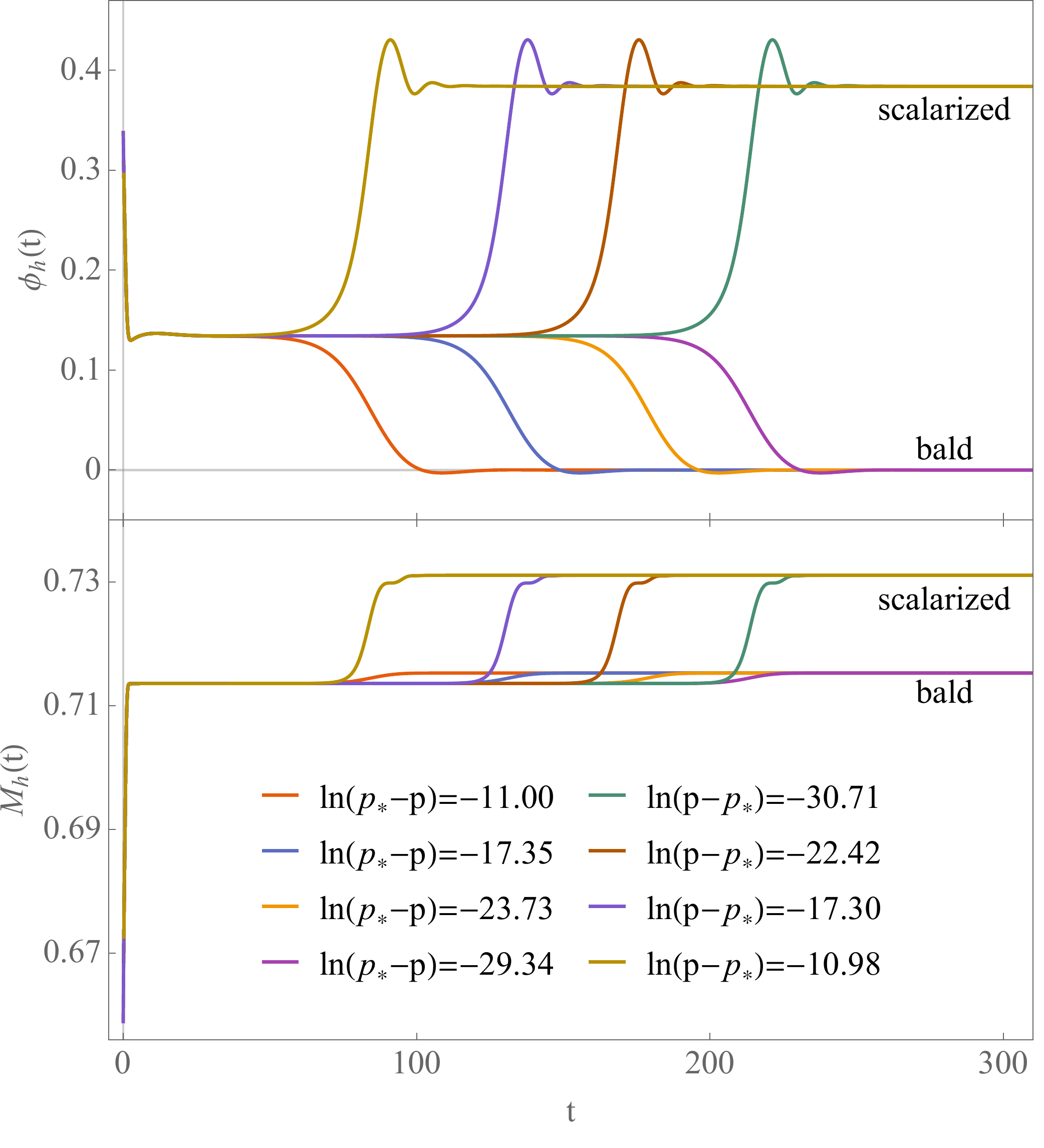}
    \par\end{centering}
    {\footnotesize{}\caption{{\footnotesize{}\label{fig:BHMphit}The evolution of 
    $\phi_h$ (upper) and 
    $M_h$ (lower) for various
    $p$ when $Q/M=0.9$ and $\beta=200$. We show the
    results obtained from initial data family $\phi=0.5e^{-(\frac{r}{pM})^{2}}$. The critical value $p_\ast \approx 2.116895853824$.
    Other initial data families lead to qualitatively the same behaviors.}}
    }{\footnotesize\par}
    \end{figure}

We fix $M=1,$ $Q/M=0.9$ and $\beta=200$ to present our results.
In  Fig.\ref{fig:BHMphit}, we show the evolution of the BH irreducible mass $M_h=\sqrt{\frac{A}{4\pi}}=r_{h}$ and the scalar value  $\phi_h$ on the apparent horizon. $A$ and $r_{h}$ denote the area and areal radius of the apparent horizon, respectively. For   initial data parameterized by $p$
(which could be $a,c$ or $w$ here), there is a threshold $p_{\ast}$. When $p$ is close to $p_\ast$, after a violent change at the very early times dominated by the initial data, all the intermediate solutions are attracted to a  CS and stay there for a long time.
By fine-tuning $p$ to the exact critical value $p_\ast$,  the evolution would stay on this CS forever, in principle. The CS corresponds to a metastable scalarized BH solution of the theory. At late times, the intermediate solutions decay to  bald RN BHs if $p<p_\ast$, or to  scalarized charged BHs if $p>p_\ast$. Note that the BH's irreducible mass never decreases with time,
satisfying the requirement of the second law. The scalar field  
displays damped oscillation at 
the late times, 
indicating that the final solutions in supercritical or subcritical cases approach linearly stable scalarized or bald BHs, respectively. 
If $p$ is far way from $p_{\ast}$, instead of being attracted to the CS, the  evolution converges directly to the  final state.

\begin{figure}[h]
    \begin{centering}
    \includegraphics[width=0.85\linewidth]{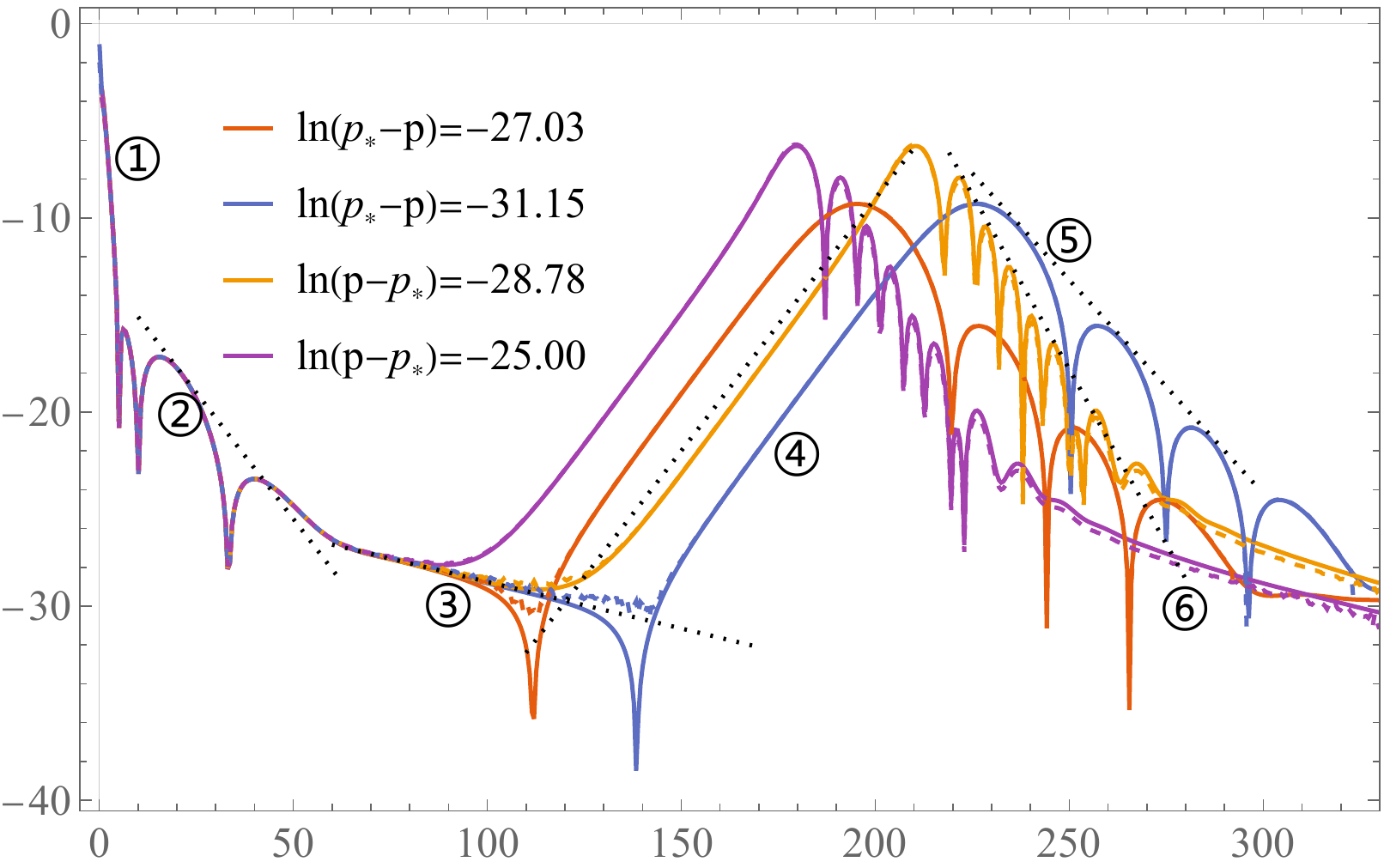}
    \par\end{centering}
    {\footnotesize{}\caption{{\footnotesize{}\label{fig:BHMphidtlog}The evolution of $1.2+2\ln|\frac{d\phi_{h}}{dt}|$ (solid) and $\ln\frac{dM_{h}}{dt}$  (dashed) for various $p$ when
    $Q/M=0.9$ and $\beta=200$. Both the evolutions can be divided into
    six stages, as labelled in the figure. Note that the evolution of $\ln\frac{dM_{h}}{dt}$  overlaps with that of $1.2+2\ln|\frac{d\phi_{h}}{dt}|$ for  almost all the stages. The dotted lines are the fitting curves for each stage.}}
    }{\footnotesize\par}
    \end{figure}

To understand the plateaus in Fig.\ref{fig:BHMphit} more clearly, we calculate  $\ln|\frac{d\phi_{h}}{dt}|$
and show the results in Fig.\ref{fig:BHMphidtlog}.  It is obvious that the nonlinear evolution can be divided
into six stages. At the first stage, the solutions are closely related
to the initial data. At the second stage, all the solutions display damped oscillation with damping rate $\nu_{\phi}\approx-0.13$. At the third stage, we get the
intermediate solutions that are very close to the critical
solution. The scalar fields  change with $|\frac{d\phi_{h}}{dt}|\propto e^{-0.024 t}$. 
At the fourth
stage, the solutions depart from the CS with $|\frac{d\phi_{h}}{dt}|\propto e^{\eta_{\phi}t}$  in which  $\eta_{\phi}\approx0.13$ for all $p$.  
At the fifth stage,
the system settles down, resembling the quasinormal
modes. The imaginary part of the dominate mode can be fitted as $\omega_{\phi I}\approx-0.18$ for the supercritical cases or $\omega_{\phi I}\approx-0.11$ for the subcritical
cases. The sixth stage is the late-time tail with $\phi_h\propto t^{-3}$ for both the supercritical and subcritical cases. 

\begin{figure}[h]
    \begin{centering}
    \includegraphics[width=0.80\linewidth]{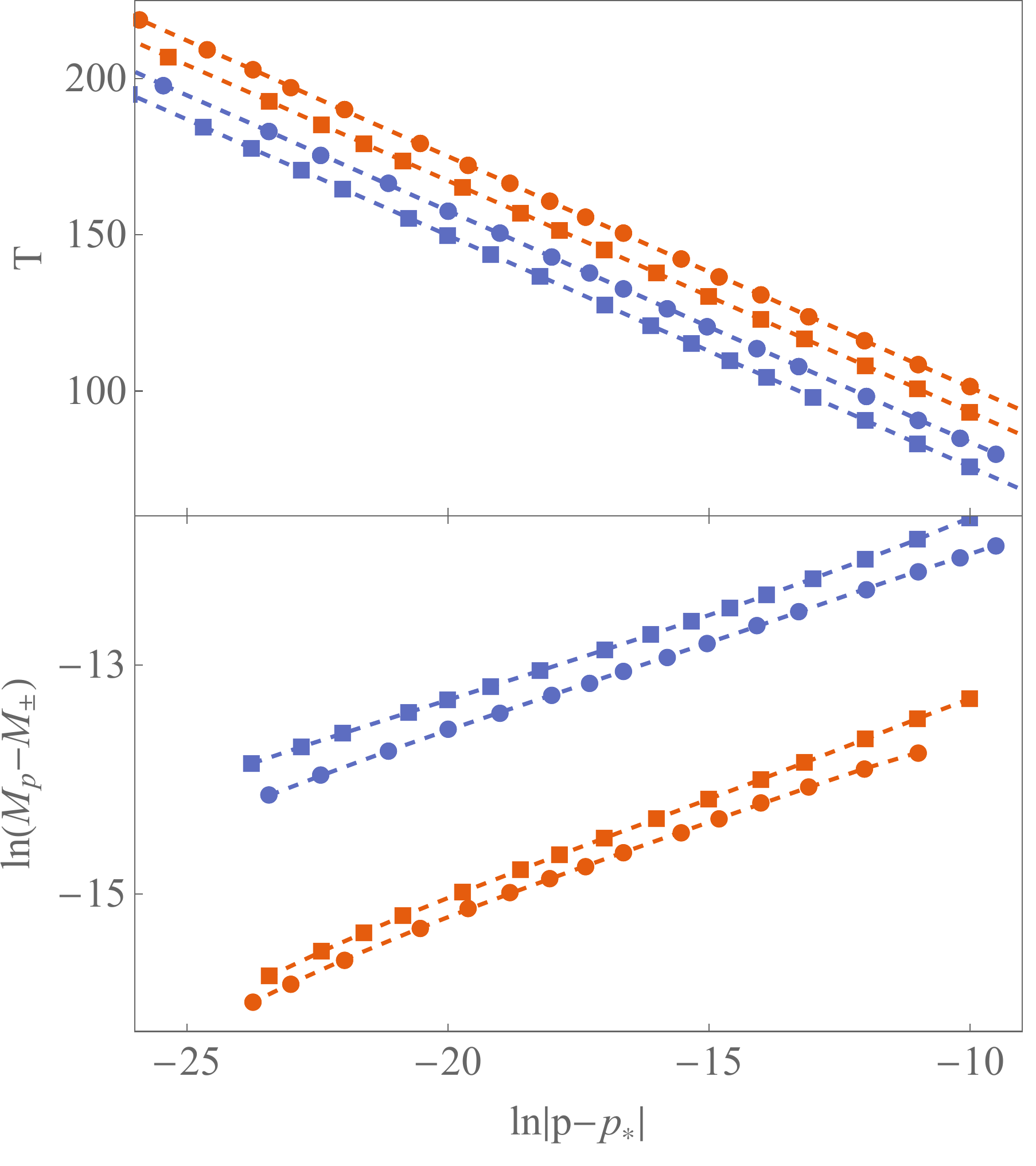}
    \par\end{centering}
    {\footnotesize{}\caption{{\footnotesize{}\label{fig:Tlog}Upper panel: The time $T$ of the
    intermediate solution that stays near the CS with respect
    to $\ln|p-p_{\ast}|$ when $Q/M=0.9$ and $\beta=200$. The lines with circle and square markers are obtained with initial conditions $\phi=0.5 e^{-(\frac{r}{pM})^{2}}$ and  $\phi=p\left(1-\tanh\frac{r}{2M}\right)$, respectively. The red lines are
    for $p<p_{\ast}$ and the blue lines for $p>p_{\ast}$.
    Here all lines have $\gamma\approx7.4$.  We use the time containing the first
    to the fourth stage as $T$, which can be get easily by counting the
    time when $\phi_{h}$ is maximum for $p>p_{\ast}$  or minimum
    for $p<p_{\ast}$, as shown in Fig.\ref{fig:BHMphit}. Since
    the first  stage takes almost the same time for
    different $p$, the coefficient $\gamma$  will not be affected. 
    Lower panel: The  power-law relations $M_p-M_{\pm}\propto|p-p_{\ast}|^{\gamma_{\pm}}$
    for $p<p_{\ast}$ (red) and $p>p_{\ast}$ (blue) for the two families of initial data.}}
    }{\footnotesize\par}
\end{figure}

It is clear that the intermediate plateaus in Fig.\ref{fig:BHMphit}
consist of the second, third and most part of the fourth stages displayed in Fig.\ref{fig:BHMphidtlog}. They are intermediate solutions $\phi_{p}(t,r)$ corresponding to $p$ that can be well approximated by  
\begin{equation}
\phi_{p}(t,r)\approx\phi_\ast(r)+(p-p_\ast)e^{\eta_\phi t}\delta \phi(r)+\text{stable modes}. \label{eq:approx}
\end{equation}
Here $\phi_\ast(r)$ stands for the critical
solution, $\delta \phi(r)$ is the only unstable eigenmode associated
with the eigenvalue $\eta_\phi$, analogues to the cases in type I critical gravitational collapse \citep{Evans:1994pj,Koike:1995jm,Bizon:1998kq}. 
At the second and third stages, the stable modes of  $\phi_{p}$ dominate. But the unstable mode  dominates in the fourth stage and grows to a finite size with time $T$ satisfying $|p-p_\ast|e^{\eta_\phi T}\sim O(1)$. For $t>T$ (the fifth and sixth stages), the solutions will be approximated by the end states instead of the CS. So $T$ is the time of intermediate solution stays near the CS. It scales as $T\propto-\gamma\ln|p-p_{\ast}|$ in which  $\gamma=\eta_\phi^{-1}\approx 7.4$ for both subcritical and supercritical cases.  This is shown in the upper panel of Fig.\ref{fig:Tlog}. 

We further show the evolution of $\ln\frac{dM_{h}}{dt}$ in Fig.\ref{fig:BHMphidtlog} which can also be divided into six stages. There is an interesting relation:
\begin{equation}
    \ln\frac{dM_{h}}{dt}\propto 2\ln|\frac{d\phi_{h}}{dt}|,\label{eq:2speed}
\end{equation}
for almost all stages. It holds not only for evolution with $p\simeq p_\ast$, but also for all initial data in our studies. This result confirm the rough relations found in \citep{Zhang:2021etr,Zhang:2021edm,Zhang:2021ybj}.
The irreducible mass equals the BH  horizon areal radius which locates at $\zeta(t,r_h)=1$. So $\frac{dr_h}{dt}=-\frac{\partial_t \zeta}{\partial_r \zeta}|_{r_h}$. Combining (\ref{eq:zr}, \ref{eq:zt}), there is $\delta r_h\sim O(\delta\phi^2)$ on the horizon for stationary solutions such as the critical or the final scalarized/bald solutions. Since the second, third and fourth stages are close to the CS, and the fifth, sixth stages are close to the final scalarized/bald BHs, (\ref{eq:2speed}) is expected to hold in these stages. But (\ref{eq:2speed}) holds even in the first stage which is highly nonlinear, so it is a rather nontrivial relation. 

In the lower panel of Fig.\ref{fig:Tlog},  we observe power-laws for the scalarized and bald BH solutions at late times:
\begin{equation}
M_{p}-M_{\pm}\propto|p-p_{\ast}|^{\gamma_{\pm}}.\label{eq:massPower}
\end{equation}
Here $M_{p},M_{\pm}$ are the irreducible masses of the final BHs  for initial data with parameters
$p$ and $p_{\pm}\to p_{\ast}$ from above/below, respectively.  Numerically, $p_\ast$
is accurate up to $10^{-12}$ due to the numerical accuracy limitation and we take $p_\pm=p_\ast\pm 10^{-12}$. The indexes $\gamma_+\approx\gamma_-\approx0.18$. The differences
$M_{p}-M_{\pm}$ result from the scalar escaping to the  infinity
during the evolution. 
The relation (\ref{eq:massPower}) is absent in type I gravitational collapse. Further,  the indexes $\gamma_{\pm}$ here are related to the families of the initial data. For example, when the initial scalar locates far away from the center, more scalar will escape to the infinity, resulting in different critical and final scalarized/bald solutions from those obtained by the initial scalar locating at the center. 
Nevertheless, we observe the power-laws for each family of the initial data. The critical indexes also depend on the coupling function \cite{Liebling:1996dx}. These   relations need more quantitative investigation in the future. 

\begin{figure}[h]
    \begin{centering}
    \includegraphics[width=0.80\linewidth]{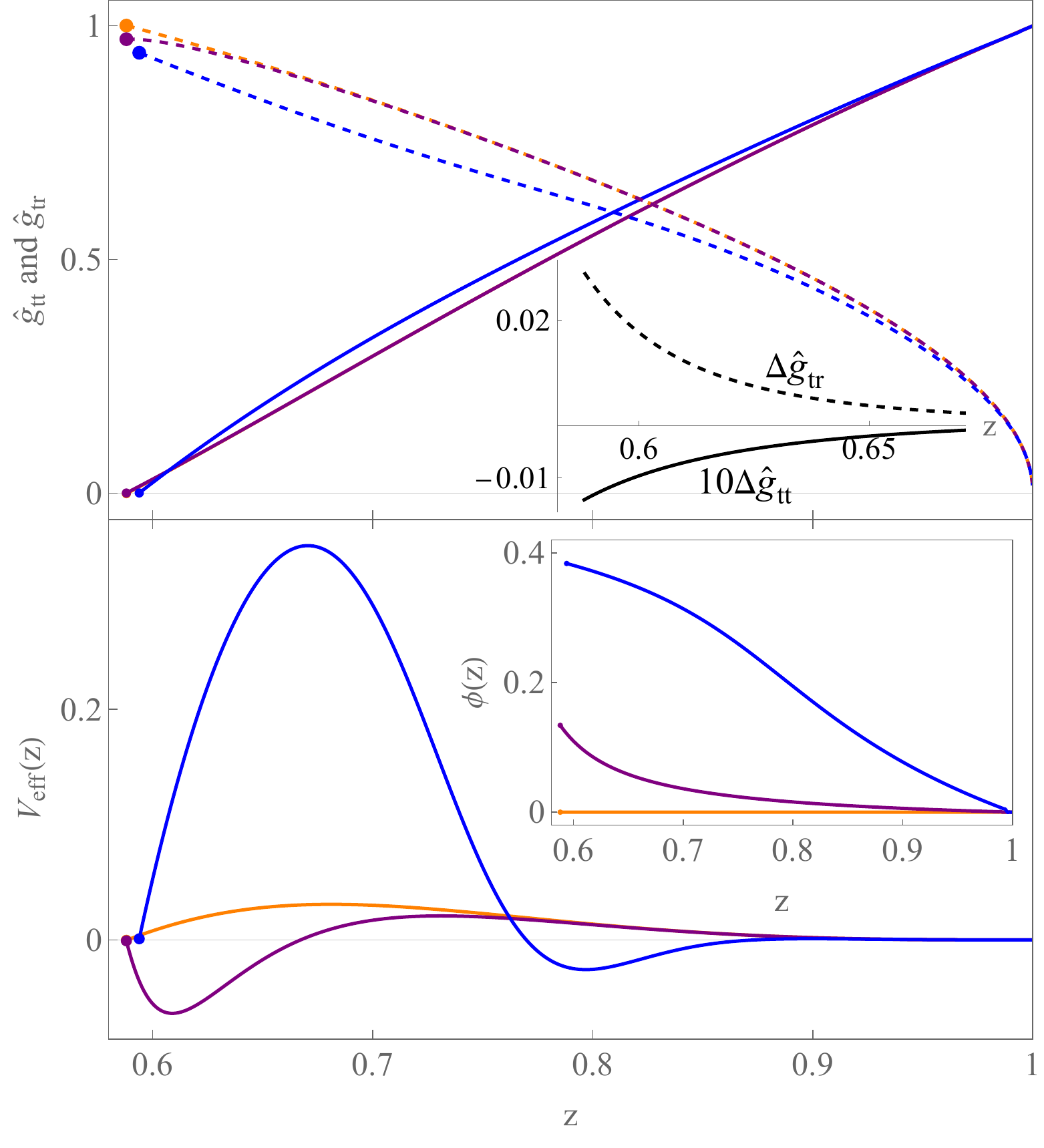}
    \par\end{centering}
    {\footnotesize{}\caption{{\footnotesize{}\label{fig:MS} Radial profiles of the metric functions $\hat{g}_{tt}=(1-\zeta^{2})\alpha^{2}$ (solid), $\hat{g}_{tr}=\zeta\alpha$ (dashed) and  $V_\text{eff},\phi$ for bald (orange), scalarized (blue) and critical (purple) solutions . 
    The upper inset shows the tiny difference between the metric functions of the bald and critical solutions.
    Here $Q/M=0.9$ and $\beta=200$. The horizons of the critical and bald/scalarized  solutions locate at $z\approx 0.5880, 0.5886,0.5938$, respectively.}}
    }{\footnotesize\par}
    \end{figure}

To understand the evolution of the CS at
late times and the linear stability of the final scalarized/bald BHs, we analyze  the radial perturbations for the critical and supercritical/subcritical solutions. 
Making a coordinate transformation $dt_{s}=dt-\zeta dr_{\ast}$ in which $dr_{\ast}=\frac{1}{(1-\zeta^{2})\alpha}dr$, supposing $\phi$ is the background scalar and $\delta\phi=e^{-i\omega t_s}\frac{R(r)}{r}$ is the  perturbation, we get a  Schr\"odinger-like equation \cite{Fernandes:2019rez}
\begin{equation}
    0=\left(\partial_{r_{\ast}}^{2}+\omega^{2}-V_{\text{eff}}\right)R.\label{eq:perturb}
\end{equation}
Here the effective potential $V_{\text{eff}}=\frac{\left(1-\zeta^{2}\right)\alpha^{2}}{r^{2}}[\zeta^{2}-2r^{2}\phi'{}^{2}-\frac{Q^{2}}{r^{2}f}(1-2r^{2}\phi'{}^{2}+\frac{2r\phi'\dot{f}}{f}+\frac{f\ddot{f}-\dot{f}{}^{2}}{2f^{2}})]$ in which $\phi'=\partial_r \phi$ and $\dot f = \frac{df}{d\phi}$. 
The distributions of the metric functions, the background $\phi$ and $V_\text{eff}$ for the critical and bald/scalarized  solutions are shown in Fig.\ref{fig:MS}. Only for the CS, 
there is  $\int_{-\infty}^\infty V_{\text{eff}}dr_\ast<0$ so that the CS cannot be stable \cite{Buell1995}. Actually, the scalar perturbation has a negative effective mass squared near the horizon. 
So the CS has tachyonic instability, which gives precisely the unstable mode in  $\phi_{p}(t,r)$
and drives the system away from  $\phi_\ast(r)$. 
As implied in  (\ref{eq:approx}), if $p<p_\ast$, the scalar surrounding the critical BH decreases to zero by falling into the BH, except a part of the scalar field disperses to the infinity, resulting in a final RN BH. 
For $p>p_\ast$, the coupling function $f(\phi)$ 
plays the role of  transforming the electromagnetic energy into the scalar field. 
The accretion of the energetic scalar field results in significant growth of the BH mass. 
On the other hand, the final scalar field  has an effective potential barrier to balance the gravity such that it survives out of the horizon.


Using the first order WKB method for the CS, we can get the dominant mode of the second stage $\omega\approx0.19-0.13i$. The dominant modes of the fifth stage are $\omega\approx 0.61 - 0.18i$ and  $\omega\approx 0.20 - 0.097i$ for the final scalarized/bald BHs, respectively. 
Note that all the imaginary part results from the linear analysis are consistent with nonlinear numerical fitting results within the error tolerance.

\section{Summary and Discussion}

We have found a new BH scalarization mechanism through the nonlinear accretion of the scalar field in EMS theories.  This mechanism is different from the previously disclosed spontaneous scalarization triggered by the linear tachyonic instability of the bald BH, in which either the bald or scalarized BH can stay stable at the same model parameter space point. We uncover a new family of critical solutions, which separate solutions containing scalarized BHs from those containing bald BHs. In particular, we have found the dynamical critical phenomena in the bald/scalarized BH phase transition, which is analogous to those observed in type I critical gravitational collapse. The discovery of this new critical behavior has revealed interesting nonlinear dynamics at the threshold of black hole scalarization and opened up a fascinating area of research in generalized theories of relativity and in the properties of spacetimes. 

\section*{Acknowledgments }
We thank  Peng-Cheng Li, Peng Liu, Chao Niu, Xiao-Ning Wu, and Hongbao Zhang for helpful discussions. This research is supported by National Key R\&D Program of China under
Grant No.2020YFC2201400, and the Natural Science Foundation of China
under Grant Nos. 11975235, 12005077 and 12035016, and Guangdong Basic and Applied Basic Research
Foundation under Grant No. 2021A1515012374. The work of B. W. was partially supported by NNSFC under grant 12075202.

\end{document}